\documentclass[aps,pra,twocolumn,showpacs,superscriptaddress,bibliography]{revtex4-1}
\usepackage{graphicx}    
\usepackage{dcolumn}    
\usepackage{bm}             
\usepackage{amssymb}   
\usepackage{amsmath}    
\usepackage{subfigure}    
\usepackage{braket}
\usepackage{mathrsfs}
\usepackage{multirow}
\usepackage[export]{adjustbox}
\usepackage[toc,page]{appendix}
\begin{document}

\title{Quantum logic for state preparation, readout, and leakage detection with binary subspace measurements}
\author{R. Tyler Sutherland}
\email{tyler.sutherland@oxionics.com}
\affiliation{Oxford Ionics Limited, Begbroke Science Park, Begbroke, OX5 1PF, UK}

\date{\today}

\begin{abstract}
We discuss a general technique for using quantum logic spectroscopy to perform quantum non-demolition (QND) measurements that determine which of two subspaces a logic ion is in. We then show how to use the scheme to perform high fidelity state preparation and measurement (SPAM) and non-destructive leakage detection, as well as how this would reduce the engineering task associated with building larger quantum computers. Using a magnetic field gradient, we can apply a geometric phase interaction to map the magnetic sensitivity of the logic into onto the spin-flip probability of a readout ion. In the `low quantization field' regime, where the Zeeman splitting of the sublevels in the hyperfine ground state manifold are approximately integer multiples of each other, we show how to we can use the technique to distinguish between $2^{K}$ hyperfine sublevels in only $K$ logic/readout measurement sequences\textemdash opening the door for efficient state preparation of the logic ion. Finally, we show how the binary nature of the scheme, as well as its potential for high QND purity, let us improve SPAM fidelities by detecting and correcting errors rather than preventing them.
\end{abstract}
\maketitle

\section{Introduction}\label{sec:intro}

Technologies like quantum computers and atomic clocks rely on quantum state preparation and measurement (SPAM) before/after a series of coherent operations in which we must prevent population from `leaking' out of a specified subspace \cite{wineland_1998, divincenzo_2000,ladd_2010,nielsen_2010,brewer_2019,an_2022,sotirova_2024}. For quantum computers to reach their industrial potential, we must maintain these capabilities as we make systems bigger \cite{pino_2021,moses_2023}, and, for clocks, as we make systems smaller and cheaper to manufacture \cite{newman_2019}. In many systems, SPAM can be challenging, slow, and error prone. Likewise, spectator states lead to the possibility of leakage, which would be detrimental to quantum error correction \cite{lidar_2013}. While solutions exist, for example by combining microwave \cite{tretiakov_2021} and/or optical quadrupole transitions with traditional optical pumping \cite{an_2022}, these issues increase the difficulty of engineering next-generation clocks and computers. \\

One solution is to employ repeated in-sequence quantum non-demolition (QND) measurements \cite{hume_2007}. For example, this method was used in Ref.~\cite{sotirova_2024}, achieving record SPAM fidelities not through improved field engineering, but rather QND measurements followed by conditional correction pulses and/or post-selection; this is particularly notable because it shows you can achieve higher fidelities through detecting and correcting SPAM errors, rather than preventing them altogether. Quantum logic spectroscopy (QLS) \cite{wineland_1998, schmidt_2005}, where information about the state of a logic ion is mapped onto the state of a readout ion, is another technique researchers have used to demonstrate high fidelity readout of trapped ion qubits \cite{hume_2007, erickson_2022}, and, with lower fidelities, to prepare/readout trapped molecular ions \cite{chou_2017}. If implemented in a quantum computing architecture based on laser-free coherent operations \cite{wineland_1998,mintert_2001,ospelkaus_2008,ospelkaus_2011,harty_2014,harty_2016,srinivas_2018,sutherland_2019,srinivas_2021,loschnauer_2024}, using QLS for SPAM could remove the need for logic ion lasers altogether; this would reduce the laser complexity associated with scaling \textit{and} decouple the choice of qubit ion from its associated lasers. Similar to Ref.~\cite{sotirova_2024}, QLS offers a way to improve SPAM fidelities without necessarily improving control field precision. At least to the extent the logic/readout sequence maintains quantum non-demolition (QND) purity, i.e. the operation does not change the logic ion's state, we can suppress SPAM errors arbitrarily through redundant operations \cite{hume_2007}\textemdash the same idea that makes modern (classical) computing possible. Experimental noise will always `break' QND purity to a certain extent, so even if a scheme is fully QND in \textit{idealized} circumstances, we must also understand the extent to which this is maintained in the presence of the largest sources of noise. To this end, Ref.~\cite{mur_2012} pointed out that the geometric phase shift interaction originally proposed in Ref.~\cite{leibfried_2003} can be combined with $\pi/2$-pulses on the readout ion to effectively map the magnetic/electric sensitivity of a molecular ion onto the rotation angle of a readout ion; measuring the readout ion's spin thus determines the shift sensitivity of the molecule, pinpointing its state without ever driving any of its transitions directly. The technique is especially promising if we generate the necessary spin-dependent force with the near-field of a magnetic gradient oscillating near the frequency of a motional mode \cite{ospelkaus_2008}, rather than a laser, since the absence of photon scattering \cite{ospelkaus_2008,ospelkaus_2011,harty_2016,weidt_2016,srinivas_2018,srinivas_2021} leaves mostly motional \cite{sutherland_2021_2} and phase shift errors \cite{srinivas_2021}\textemdash neither of which violate the QND purity of the operation. \\

In this work, we discuss how to use the general interaction proposed in Ref.~\cite{mur_2012} to implement efficient, high-fidelity SPAM (Sec.~\ref{sec:spam}), and non-destructive leakage detection (Sec.~\ref{sec:leakage}). Focusing on the `low quantization field' regime, i.e. when the Zeeman splittings of the $S_{1/2}$ ground state manifold are (nearly) integer multiples of one another, we discuss how to engineer logic/readout cycles that map the subspace of potential logic ion states onto two orthogonal states of the readout ion. This allows us to eliminate up to \textit{half} of the logic ion's hyperfine sublevels with each logic/readout cycle, regardless of measurement outcome. This lets us distinguish between up to $2^{K}$ states in only $K$ logic/readout sequences, in ideal circumstances; for example, in Sec.~\ref{sec:preparation} we discuss how to do state preparation on the $4$ hyperfine sublevels of $^{171}\text{Yb}^{+}$ in $K=2$ measurement cycles, and the $8$ hyperfine sublevels of $^{137}\text{Ba}^{+}$ in $K=3$ measurement cycles (see appendix). In the presence of noise, we discuss how to use classical error correction techniques to suppress subspace measurement errors. Finally, we discuss how to leverage additional subspace measurements to suppress errors from the shelving pulses necessary for the scheme\textemdash even though shelving pulses themselves do not preserve QND purity.

\section{State-dependent rotations}\label{sec:sdr}

We consider a trapped ion mixed-species crystal comprising a `logic' and a `readout' ion. We assume the logic ion has non-zero nuclear-spin $I\neq 0$, and focus on the hyperfine sublevels of the $S_{1/2}$ ground state. We represent the logic ion's interaction with magnetic fields using the spin-angular momentum operators $\vec{J}$. For simplicity, we assume the readout ion has zero nuclear spin, and therefore only two ground states; we label these states $\ket{0}$ and $\ket{1}$, and represent its field interaction with standard Pauli operators $\vec{\sigma}$. We take the direction of the quantization field $B_{q}$ to be $z$, making the total Hamiltonian:
\begin{eqnarray}\label{eq:hyperfine_zeeman_orig}
    \hat{H}_{q}=\hbar A \vec{J}^{\prime}\cdot \vec{I} + \mu_{B}g_{J}B_{q}\Big(\hat{J}_{z}+\frac{1}{2}\hat{\sigma}_{z}\Big),
\end{eqnarray}
where $A$ is the hyperfine constant for the logic ion \cite{langer_2005}, and we have ignored the (weak) interaction of the nucleus with $B_{q}$. We describe the logic ion hyperfine sublevels using $\ket{F^{\pm},m_{F}}$, where $F^{\pm}\equiv I\pm 1/2$. Since it has no hyperfine structure, $\hat{H}_{q}$ is already diagonal with respect to the readout ion, and we can diagonalize the logic ion subspace exactly using the Breit-Rabi formula \cite{breit_1931, langer_2005}, which we represent with a unitary operator $\hat{U}_{B}$. In the following, we assume this basis, and deal only with transformed spin-angular momentum operators:
\begin{eqnarray}
    \vec{J} &\rightarrow & \hat{U}^{\dagger}_{B}\vec{J}^{\prime}\hat{U}_{B},
\end{eqnarray}
which we will tacitly assume throughout the remainder of this work. We focus on the `low quantization field' regime $B_{q}\lesssim 10~\text{G}$, where $\hbar A \gg \mu_{G}g_{J}B_{q}$. As a result, the frequency of each hyperfine sublevel versus the magnetic field projected onto the quantization axis $B_{z}$ is approximately linear; specifically, $\bra{F^{\pm},m_{F}}\hat{J}_{z}\ket{F^{\pm},m_{F}}\equiv b_{F^{\pm},m_{F}}\rightarrow \pm m_{F}/2 F^{+}$ in the limit $B_{q}\rightarrow 0$. \\

If, in addition to $B_{q}$, both ions experience a sinusoidally oscillating magnetic field gradient along the quantization axis $B_{z}^{\prime}$ \cite{ospelkaus_2008} we can represent this as:
\begin{eqnarray}\label{eq:gate_pre_rwa}
    \hat{H}_{g} &=& \mu_{B}g_{J}B^{\prime}_{z}\cos(\omega_{g} t)\Big(\hat{J}_{z}x_{l}+\frac{1}{2}\hat{\sigma}_{z}x_{r}\Big),
\end{eqnarray}
where we assume that the gradient frequency $\omega_{g}\sim\text{MHz}$ is $\delta \sim \text{kHz}$ detuned from the gate mode frequency $\omega_{m}\sim\text{MHz}$, i.e. $\omega_{g}=\omega_{m}-\delta$, and $x_{l(r)}$ is the logic(readout) ion's position. Rewriting the motional component of this as eigenmodes, and making the rotating wave approximation, gives:
\begin{eqnarray}\label{eq:simple_gp_gate}
    \hat{H}_{g} &\simeq & \frac{\hbar\Omega_{g}}{2}\Big(\hat{J}_{z}c_{l}+\hat{\sigma}_{z}c_{r} \Big)\Big(\hat{a}^{\dagger}e^{i\delta t}+\hat{a}e^{-i\delta t} \Big),
\end{eqnarray}
where we have introduced ladder operators $\hat{a}^{(\dagger)}$, the gate Rabi frequency:
\begin{eqnarray}
    \Omega_{g}\equiv \mu_{B}g_{J}B^{\prime}_{z}\tilde{x}_{g},
\end{eqnarray}
the projection of the logic(readout) ion onto the gate mode $c_{l(r)}$, and its spatial extent $\tilde{x}_{g}$. We analyze the dynamics of $\hat{H}_{g}$ using the Magnus expansion \cite{magnus_1954} up to $2^{\text{nd}}$-order:
\begin{eqnarray}
    \hat{U}_{g}=\exp\!\Big(\!\!-\frac{i}{\hbar}\int^{t}_{0}\!\!dt^{\prime}\hat{H}_{g}(t^{\prime})-\frac{1}{2\hbar^{2}}\!\!\int^{t}_{0}\!\!\int^{t^{\prime}}_{0}\!\!\!\!dt^{\prime}dt^{\prime\prime}\Big[\hat{H}_{g}(t^{\prime}),\hat{H}_{g}(t^{\prime\prime}) \Big]\Big), \nonumber \\
\end{eqnarray}
which is exact for this $\hat{H}_{g}$ \cite{roos_2008}. We ensure the $1^{\text{st}}$-order terms vanish at $t_{g}$, the end of the operation, by setting $t_{g}=2\pi/\delta$, which gives:
\begin{eqnarray}\label{eq:magnus_phase_not_dropped}
    \hat{U}_{g}&=&\exp\!\Big(\frac{i\Omega_{g}^{2}t_{g}}{4\delta}\hat{J}_{z}^{2}c_{l}^{2}\Big)\exp\Big(\frac{i\Omega_{g}^{2}t_{g}}{2\delta}\hat{J}_{z}\hat{\sigma}_{z}c_{l}c_{r}\Big),
\end{eqnarray}
where the first term produces a phase shift on the logic ion. Below, we ignore this term because we are interested in schemes where the success/failure either depends on probabilities, rather than amplitudes (Secs.~\ref{sec:readout}-\ref{sec:preparation}), or where the qubit is magnetically insensitive (Sec.~\ref{sec:leakage}), giving:
\begin{eqnarray}
    \hat{U}_{g}&\simeq &\exp\Big(\frac{i\Omega_{g}^{2}t_{g}}{2\delta}\hat{J}_{z}\hat{\sigma}_{z}c_{l}c_{r}\Big).
\end{eqnarray}
If, before/after we generate $\hat{U}_{g}$, we apply an oscillatory magnetic field at the frequency of the readout ion transition $\omega_{r}$ to implement a $\pi/2$-pulse:
\begin{eqnarray}
    \hat{T}=\exp\Big(-\frac{i\pi}{4}\sigma_{x}\Big),
\end{eqnarray}
the total time-propagator for the system becomes:
\begin{eqnarray}\label{eq:spin_dep_rot_pre_proj}
    \hat{U}_{g}^{\prime}&=& \hat{T}^{\dagger}\hat{U}_{g}\hat{T} \nonumber \\
    &=&\exp\Big(i d\theta F^{+}\hat{J}_{z}\hat{\sigma}_{y}\Big),
\end{eqnarray}
where we introduced:
\begin{eqnarray}
    d\theta\equiv \frac{\Omega_{g}^{2}c_{l}c_{r}t_{g}}{2\delta F^{+}},
\end{eqnarray}
the difference in phase between adjacent $\Delta m_{F}=\pm 1$ sublevels. Similar to traditional geometric phase gates \cite{molmer_1999,molmer_2000,leibfried_2003}, tuning the values of $\Omega_{g}$, $\delta$, and $t_{g}$ lets us specify $d\theta$\textemdash the entanglement angle of the gate. Since the only off-diagonal terms in $\hat{J}_{z}$ are $\Delta m_{F}=0$ transitions across the hyperfine manifold ($\sim\text{GHz}$), we can take $J_{z}$ to be diagonal. Introducing $b_{F,m_{F}}\equiv \bra{F,m_{F}}\hat{J}_{z}\ket{F,m_{F}}$, we can project it onto the logic ion's state $\ket{F,m_{F}}$, giving:
\begin{eqnarray}\label{eq:spin_dep_rot_pre_proj}
    \hat{U}_{g}^{\prime}(d\theta)\!\!&=& \!\!\exp\!\Big(i d\theta F^{+}\hat{\sigma}_{y}\!\!\!\sum_{F,m_{F}}\!\!b_{F,m_{F}}\!\ket{F,m_{F}}\!\bra{F,m_{F}}\Big)  \\
    &\simeq & \!\!\exp\!\Big(\frac{i d\theta\hat{\sigma}_{y}}{2}\!\!\!\sum_{F^{\pm},m_{F}}\!\!\pm m_{F}\!\ket{F^{\pm},m_{F}}\!\bra{F^{\pm},m_{F}}\Big), \nonumber
\end{eqnarray}
as $B_{q}\rightarrow 0$. Since $\hat{U}_{g}^{\prime}$ is diagonal with respect to the logic ion subspace, we can simplify our analysis by projecting the system onto the basis states of the logic ion. If we include the possibility of an additional $y$-rotation $\phi_{y}$ on the readout ion, the total propagator for each $\ket{F,m_{F}}$ is:
\begin{eqnarray}\label{eq:final_prop_U_l}
    \hat{U}_{l}(d\theta,\phi_{y} ) &=& \exp\Big(\frac{i \theta_{F,m_{F}}}{2}\hat{\sigma}_{y}\Big),
\end{eqnarray}
where the total rotation angle is:
\begin{eqnarray}
    \theta_{F^{\pm},m_{F}}\equiv \phi_{y} \pm d\theta m_{F}.
\end{eqnarray}
This makes the spin-flip probability of the readout ion:
\begin{eqnarray}
     P_{F,m_{F}} = \sin^{2}\Big(\frac{\theta_{F,m_{F}}}{2}\Big).
\end{eqnarray}\\

\section{Subspace measurements}\label{sec:subspace_measurements}

Assume we want to measure the state of some logic ion, which initially contains non-zero population only in states $\ket{F,m_{F}}\in C$. For the discussions below, we prepare the readout ion in $\ket{0}$ at the beginning of the experiment and immediately after any measurement. For pure states, this makes the total wave function:
\begin{eqnarray}\label{eq:init_sub_measure}
    \ket{\psi_{C}} \!\!&=&\!\! \sum_{C}c_{F,m_{F}}\ket{F,m_{F}} \ket{0}.
\end{eqnarray}
At every step of our measurement sequence, our goal is to tune the values of $d\theta$ and $\phi_{y}$ such that the spin-flip probability of the readout ion is $P_{F,m_{F}}\in \{0,1\}$ for every state in $C$. For the first logic/readout cycle of the sequence, this will give:
\begin{eqnarray}
    \hat{U}_{l}\ket{\psi} &=& \sum_{A}c_{F,m_{F}}\ket{F,m_{F}} \ket{0} + \sum_{B}c_{F,m_{F}}\ket{F,m_{F}}\ket{1}, \nonumber \\
\end{eqnarray}
where $A(B)$ is the subspace of states such that the readout ion is mapped onto $\ket{0}(\ket{1})$; note that $A+B=C$. The probability of measuring $\ket{0}(\ket{1})$ is:
\begin{eqnarray}
    P_{A(B)} &=& \sum_{A(B)}|c_{F,m_{F}}|^{2},
\end{eqnarray}
which projects the system onto the state \cite{molmer_1993}:
\begin{eqnarray}
\ket{\psi_{A(B)}}&=& P_{A(B)}^{-1/2}\sum_{A(B)}c_{F,m_{F}}\ket{F,m_{F}}\ket{0}.
\end{eqnarray}
Conditioned on the outcome of this measurement, we may or may not apply shelving $\pi$-pulses to the logic ion. 

\subsection{SPAM protocol}\label{sec:spam}

If the initial state of the system $\ket{\psi_{0}}$ has more than two sublevels with non-zero population $\ket{F,m_{F}}\in C_{0}$, we need multiple logic/readout cycles to distinguish every state. Let $C_{j}$ be subspace of states after the $j^{th}$ logic/readout cycle, and $P_{C_{j}}$ be the probability that we made the series of measurements that project the logic ion onto $C_{j}$. If the next logic/readout cycle maps the system onto $A_{j}(B_{j})$ such that $A_{j} + B_{j}=C_{j}$, the probability of first measuring $C_{j}$ then $A_{j}(B_{j})$ is:
\begin{eqnarray}
    P_{C_{j},A_{j}(B_{j})} &= & P_{C_{j}}P_{A_{j}(B_{j})} \nonumber \\
    &=& \sum_{A_{j}(B_{j})}|c_{F,m_{F}}|^{2},
\end{eqnarray}
equal to the probability we would have measured the logic ion's state $\ket{F,m_{F}}\in A_{j}(B_{j})$ by directly measuring $\ket{\psi_{0}}$. For the next logic/readout cycle, this makes $C_{j+1}\equiv A_{j}(B_{j})$. We can thus repeat $K$ cycles until the final subspace $C_{K}$ contains a single pure state $\ket{F,m_{F}}$ with probability:
\begin{eqnarray}
    P_{F,m_{F}}&=&|c_{F,m_{F}}|^{2}.
\end{eqnarray}
We can use this protocol to readout every state in $C_{0}$ so long as every measurement at step $j$ eliminates at least one state from $C_{j}$, for every possible $C_{j}$. Because there is no restriction that $C_{1}$ comprises two states, we can use the technique to readout qubits and qudits, though our examples focus on the former. As we discuss in Sec.~\ref{sec:preparation}, it is possible to construct readout sequences that eliminate half the states in $C_{j}$ for every measurement in the sequence\textemdash letting us measure a set of $2^{K}$ states in just $K$ cycles. \\

This protocol straightforwardly generalizes to mixed states, as do all the use cases below. To see this, note that a mixed state is equivalent to a system with probability $P_{j}$ of being in the pure state $\ket{\psi_{j}}$ and that the above scheme maps \textit{any} pure state $\ket{\psi}\in C_{0}$ onto a known pure state. We can, therefore, use similar protocols for SPAM. Assuming the initial state of the logic ion is a stochastic mixture of every $\ket{F,m_{F}}$, we can interpret state preparation as the special case of a general protocol for measuring $C_{0}$, such that $C_{0}$ comprises every sublevel in the $S_{1/2}$ ground state manifold.

\subsection{Qubit leakage detection}\label{sec:leakage}

\begin{figure}[b]
\includegraphics[width=0.5\textwidth]{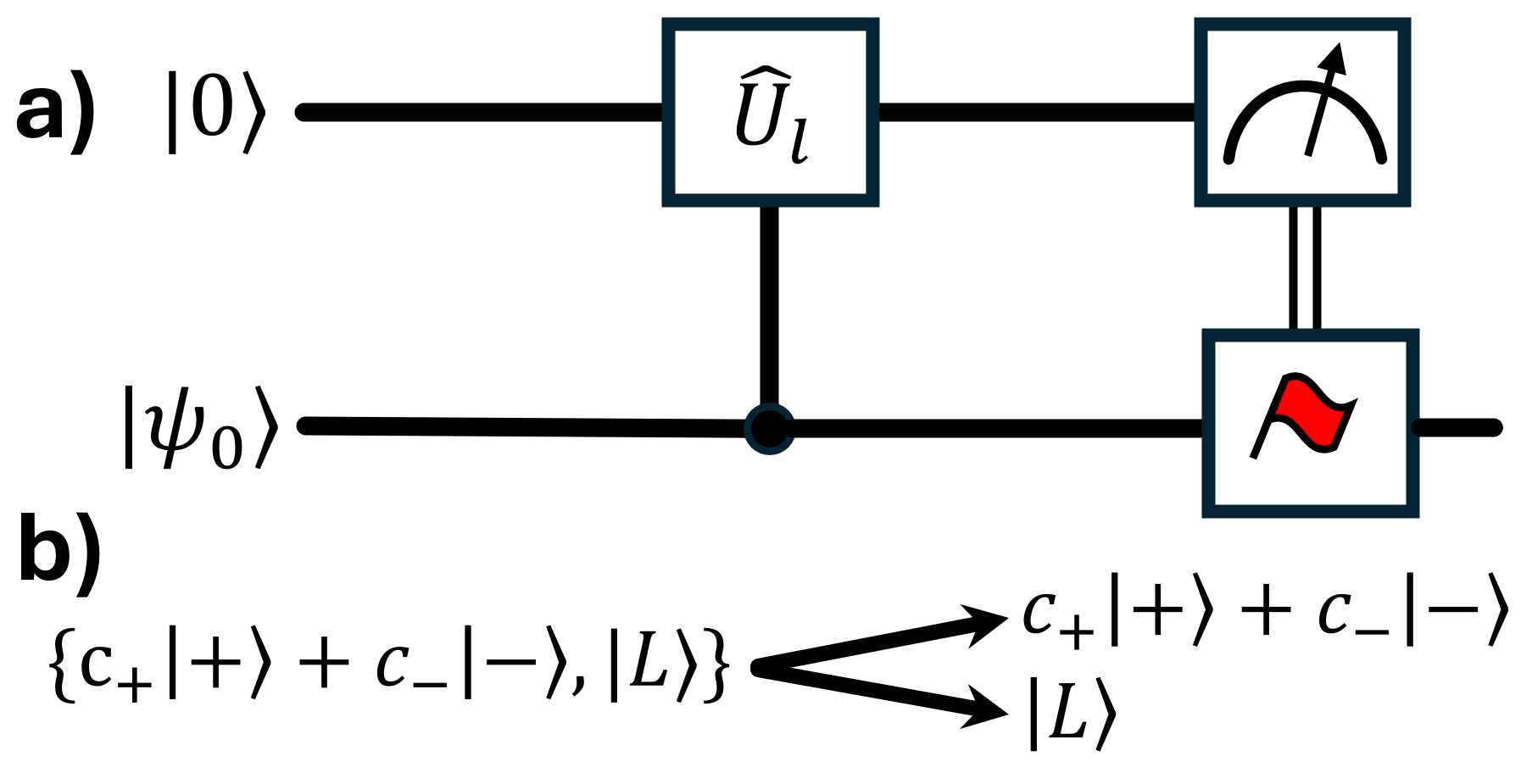}
\centering
\caption{a) Circuit elements for non-destructive leakage detection and b) our knowledge of the logic ion's state during the circuit. First, we apply the conditional rotation operation $\hat{U}_{l}$ to a magnetically insensitive qubit. If the logic ion is in the qubit manifold, $\hat{U}_{l}$ leaves the readout ion unchanged. If, on the other hand, population leaked to some $\Delta m_{F}=\pm 1$ state $\ket{L}$, then $\hat{U}_{l}$ flips the readout ion to $\ket{1}$. Measuring $\ket{0}$ projects the system onto a state where leakage has not occurred, while measuring $\ket{1}$ flags an error.}
\label{fig:leakage_circuit}
\end{figure}

Consider a system where the logic ion's population is entirely contained within a magnetically insensitive qubit subspace $Q$ \cite{langer_2005}. This means that $\hat{U}_{l}$ will induce the same rotation angle $d\theta$ on the readout ion for both states $\ket{F,m_{F}}\in Q$. In this case, measuring the readout ion gives no information about the qubit's state and projects the logic ion onto $Q$. Measuring a different rotation angle, however, flags a leakage error and projects the system onto a subspace $L\notin Q$. For example, let $Q\equiv \{\ket{-},\ket{+}\}$ for some pair of magnetically insensitive sublevels, and $\ket{L}$ be a some leakage state $\Delta m_{F}=\pm 1$. If a small amount of population has leaked to $\ket{L}$, the total wave function is:
\begin{eqnarray}
    \ket{\psi_{0}}\ket{0} &=& \Big(c_{+}\ket{+}+c_{-}\ket{-} +\varepsilon\ket{L}\Big)\ket{0},
\end{eqnarray}
where $|\varepsilon| \ll |c_{\pm}|$. Applying $\hat{U}_{l}$ with $d\theta$ and $\phi_{y}$ such that the qubit states do not flip the readout ion's spin, but $\ket{L}$ does, gives:
\begin{eqnarray}
    \hat{U}_{l}\ket{\psi_{0}}\ket{0}&=& \Big(c_{+}\ket{+}+c_{-}\ket{-}\Big)\ket{0} +\varepsilon\ket{L}\ket{1}.
\end{eqnarray}
If we measure the readout ion's spin as $\ket{0}$, the system is projected onto:
\begin{eqnarray}
    \ket{\psi_{A}} &=& \Big(c_{+}\ket{+}+c_{-}\ket{-}\Big)\ket{0},
\end{eqnarray}
leaving the qubit unperturbed, while measuring $\ket{1}$ flags a leakage error and projects the system onto:
\begin{eqnarray}
    \ket{\psi_{B}} &=& \ket{L}\ket{1}.
\end{eqnarray}
In Fig.~\ref{fig:leakage_circuit}a, we provide a circuit illustration of this sequence and, in Fig.~\ref{fig:leakage_circuit}b, we indicate our knowledge of the logic ion's state during the sequence. Finally, this protocol does not rely on the phase of $\varepsilon$, and can be used to flag incoherent leakage as well.

\section{Example implementations}
\begin{figure}[b]
\includegraphics[width=0.45\textwidth]{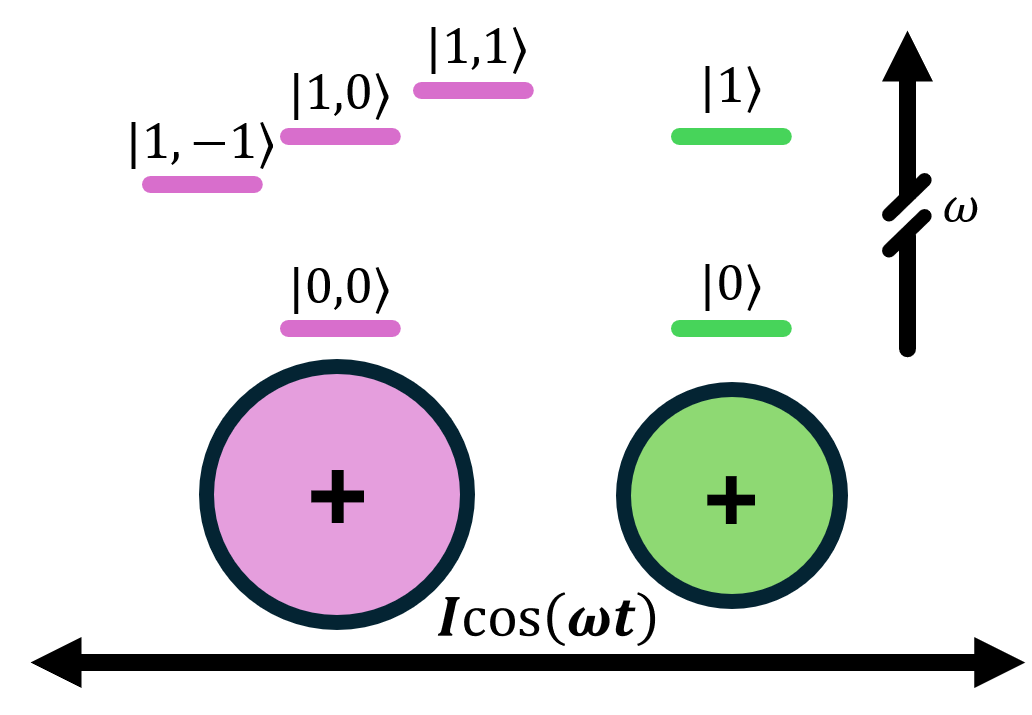}
\centering
\caption{We consider a mixed-species two ion crystal comprising a nuclear spin $I \neq 0$ logic ion and two states $\{\ket{0},\ket{1}\}$ of a readout ion. In line with examples in the main text, we show a $I=1/2$ $S_{1/2}$ ground state manifold corresponding to $^{171}\text{Yb}^{+}$. We generate a $\sim\text{MHz}$ frequency gradient, tuned to the frequency of one of the crystal's motional modes to map information from the logic ion to the readout ion, while preserving QND purity. This implements a geometric phase gate and, 2hen combined with $\pi/2$-pulses on the readout ion, induces a spin-dependent rotation on the readout ion.}
\label{fig:figure_1}
\end{figure}
\subsection{Qubit Readout of $^{171}\text{Yb}^{+}$}\label{sec:readout}

Typically, trapped ion systems store information in qubit subspaces that are insensitive to slow moving magnetic fields, so we need to shelve the qubit states before we can distinguish them. For this example, we take the qubit to be $\{\ket{1,0},\ket{0,0}\}$ of the $^{171}\text{Yb}^{+}$ ground state manifold:
\begin{eqnarray}
    \ket{\psi_{0}} &=& c_{0,0}\ket{0,0}+c_{1,0}\ket{1,0}.
\end{eqnarray}
This is magnetically insensitive, so we must first apply $\pi$-pulses to shelve $\ket{0,0}\rightarrow \ket{1,1}$ then $\ket{1,0}\rightarrow \ket{0,0}\rightarrow \ket{1,-1}$. The logic ion state is then:
\begin{eqnarray}
    \hat{U}_{s}\ket{\psi_{0}} &=& c_{0,0}\ket{1,1}+c_{1,0}\ket{1,-1},
\end{eqnarray}
up to a phase. Assuming the low quantization field regime, $B_{q}\lesssim 10~\text{G}$, the rotation angle of each state is:
\begin{eqnarray}
\theta_{F^{\pm},m_{F}} &\simeq & \phi_{y} \pm d\theta m_{F}.
\end{eqnarray}
If we set $d\theta=\pi/2$ and $\phi_{y}=\pi/2$, applying $\hat{U}_{l}$ gives:
\begin{eqnarray}
    \hat{U}_{l}\hat{U}_{s}\ket{\psi_{0}}\ket{0} &=& c_{0,0}\ket{1,1}\ket{0}+ c_{1,0}\ket{1,-1}\ket{1}. \nonumber 
\end{eqnarray}
After this, the probability we measure the readout ion as $\ket{0(1)}$ is $P_{0(1)}=|c_{0(1),0}|^{2}$. \\

\subsection{Initialization of $^{171}\text{Yb}^{+}$}\label{sec:preparation}

\begin{figure}[b]
\includegraphics[width=0.5\textwidth]{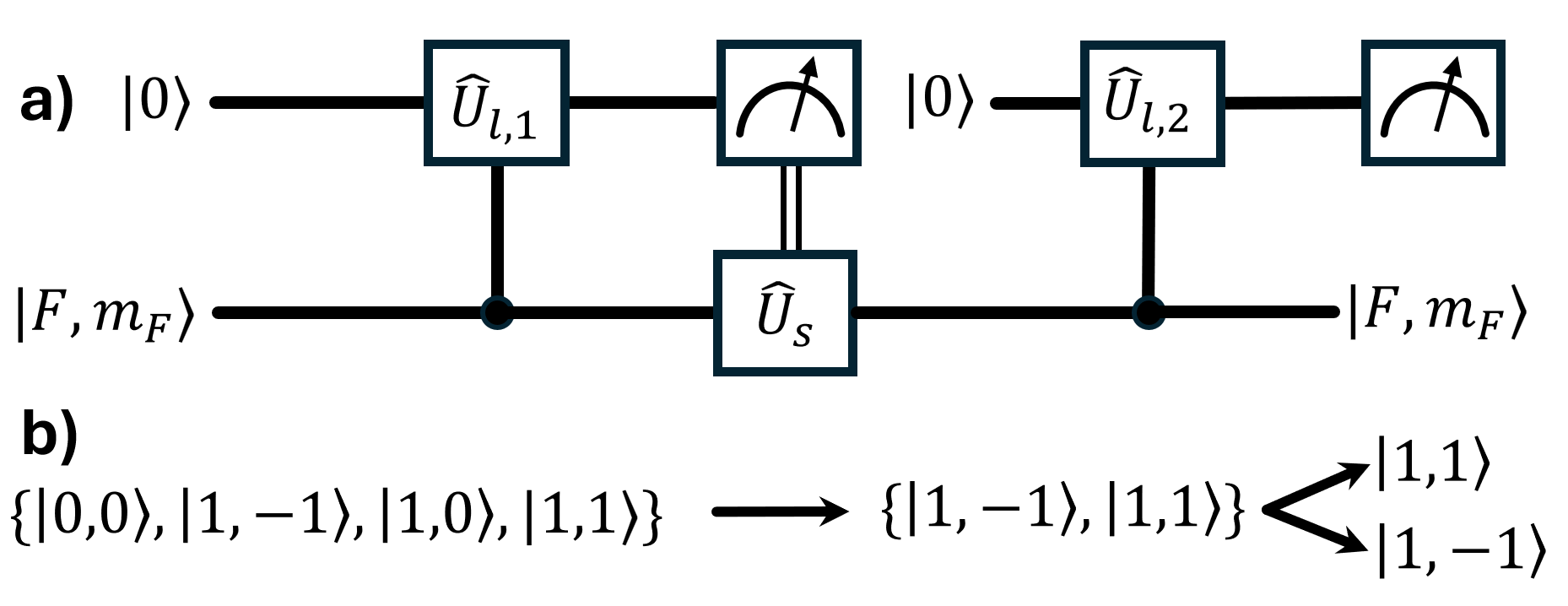}
\centering
\caption{a) Circuit for $^{171}\text{Yb}^{+}$ state preparation scheme, and b) our knowledge of the logic ion's state at each stage of the process. First, we apply our conditional rotation operation with $\hat{U}_{l}$, setting $d\theta =\pi$ and $\phi_{y}=0$. If we measure the readout ion as $\ket{1}$, we know the state is in the $\{ \ket{1,-1},\ket{1,1}\}$ subspace. If we measure the readout ion's state as $\ket{0}$, we apply our conditional shelving pulses to map $\{\ket{0,0},\ket{1,0}\}\rightarrow \{\ket{1,-1},\ket{1,1}\}$. Finally, we apply a second $\hat{U}_{l}$, setting $d\theta =\pi/2$ and $\phi_{y}=\pi/2$, after which measuring the readout ion will identify the logic ion's state.} 
\label{fig:yb_init}
\end{figure}

To initialize $^{171}\text{Yb}^{+}$, we need to distinguish between all four sublevels of the $S_{1/2}$ manifold, making $C_{0}\equiv \{\ket{0,0},\ket{1,-1},\ket{1,0},\ket{1,1} \}$. For the first logical operation $\hat{U}_{l,1}$, we set $d\theta = \pi$ and $\phi_{y}=0$. Applying $\hat{U}_{l,1}$ gives:
\begin{eqnarray}
    \hat{U}_{l,1}\ket{F,m_{F}}\!\ket{0}\!\!\! &=&\! \!\!\ket{F,m_{F}}\Big(\!\!\cos\!\Big[\frac{m_{F}\pi}{2} \Big]\!\!\ket{0}\!\pm i\sin\!\Big[\frac{m_{F}\pi}{2} \Big]\!\ket{1}\!\Big), \nonumber \\
\end{eqnarray}
indicating the readout ion is in the state $\ket{0}$ if $m_{F}$ is even, and $\ket{1}$ if $m_{F}$ is odd. Measuring the readout ion's state as $\ket{0}$ or $\ket{1}$, therefore, tells us if the logic ion is in subspace $A_{1}\equiv \{\ket{0,0},\ket{1,0} \}$ or $B_{1}\equiv \{\ket{1,-1},\ket{1,1} \}$, respectively. If we are in $A_{1}$, we need to apply specific shelving pulses $\hat{U}_{s}$ to drive $\ket{0,0}\rightarrow\ket{1,1}$ and then $\ket{1,0}\rightarrow\ket{0,0}\rightarrow\ket{1,-1}$; this maps $A_{1}\rightarrow B_{1}$, so $C_{1}\equiv B_{1}$ regardless of the initial measurement value. For the second logical operation $\hat{U}_{l,2}$, we set $d\theta=\pi/2$ and $\phi_{y}=\pi/2$, which maps:
\begin{eqnarray}\label{eq:conditional_mapping}
    \hat{U}_{l,1}\ket{1,1}\ket{0} &\rightarrow & \ket{1,1}\ket{1} \nonumber \\
    \hat{U}_{l,1}\ket{1,-1}\ket{0} &\rightarrow & \ket{1,-1}\ket{0},
\end{eqnarray}
after which measuring the readout ion identifies the qubit state\textemdash identifying the state in $K=2$ logic/readout sequences.

\subsection{Leakage detection in $^{171}\text{Yb}^{+}$}
Consider a $^{171}\text{Yb}^{+}$ atom with most of it's population in the magnetically insensitive $Q\equiv \{\ket{1,0},\ket{0,0}\}$ qubit; because of the quadratic coupling between the hyperfine manifolds, this qubit will not actually be magnetically insensitive at non-zero $B_{q}$, but the resulting phase shift from the $\propto \hat{J}_{z}^{2}$ in Eq.~(\ref{eq:magnus_phase_not_dropped}) can be tracked \cite{pino_2021,moses_2023}, and, as we discuss in Sec.~\ref{sec:infidelity_redundancy}, we can classically correct measurement errors. If we have a qubit manifold with non-zero population that has leaked out of $Q$, the total state of the system is:
\begin{eqnarray}
    \ket{\psi_{q}}\ket{0}&=& \Big(a\ket{1,0} + b\ket{0,0} + \alpha\ket{1,-1} + \beta\ket{1,1}\Big)\ket{0}, \nonumber \\
\end{eqnarray}
where $\alpha,\beta \ll a, b$. Like we found in the previous section, applying $\hat{U}_{l}(\pi,0)$ to the entire $^{171}\text{Yb}^{+}$ hyperfine ground state manifold maps the readout ion onto $\ket{0}$ if the logic ion is in $Q$ and $\ket{1}$ if not:
\begin{eqnarray}
    \hat{U}_{l}(\pi,\!0)\ket{\psi_{q}}\!\ket{0}\!\!\!&=&\!\!\! \Big(\!a\ket{1,0} \!+ \!b\ket{0,0}\!\Big)\ket{0}\! + \!\Big(\!\alpha\ket{1,-1} \!+ \!\beta\ket{1,1}\!\Big)\!\ket{1}. \nonumber \\ 
\end{eqnarray}
Measuring the readout ion's state as $\ket{0}$, i.e. that no leakage has occurred, the system experiences a quantum jump \cite{molmer_1993} that projects the system onto the $\ket{0}$ state of the readout ion:
\begin{eqnarray}
    \ket{0}\bra{0}\hat{U}_{l}(\pi,\!0)\ket{\psi_{q}}\!\ket{0}\!\!\!&=&\!\!\! \Big(\!a\ket{1,0} \!+ \!b\ket{0,0}\!\Big)\ket{0}.
\end{eqnarray}
Up to a normalization factor, this leaves the qubit manifold unchanged, while also eliminating the population associated with the `leaked' states. If we measure $\ket{1}$, on the other hand, the system is projected onto:
\begin{eqnarray}
    \ket{1}\bra{1}\hat{U}_{l}(\pi,\!0)\ket{\psi_{q}}\!\ket{0}\!\!\!&=&\!\!\! \Big(\!\alpha \ket{1,-1} \!+ \!\beta \ket{0,0}\!\Big)\ket{1},
\end{eqnarray}
destroying any information held in the qubit subspace and flagging a leak. 

\section{Error correction}\label{sec:infidelity_redundancy}

In this section, we discuss how to modify the measurement techniques described above to suppress any errors that do not change the state of the logic ion, i.e. violate QND purity. To understand this, suppose we apply some Hamiltonian $\hat{H}_{0}$ to measure a system observable $\hat{O}$. If at all times:
\begin{eqnarray}\label{eq:qnd_def}
    \Big[\hat{H}_{0},\hat{O}\Big] = 0
\end{eqnarray}
this is a quantum non-demolition QND measurement \cite{caves_1980,meunier_2006,lupacscu_2007}; here $\hat{O}\equiv \ket{F,m_{F}}\bra{F,m_{F}}$, since we wish to measure the logic ion's state. To the extent Eq.~(\ref{eq:qnd_def}) remains valid in the presence of noise, control or environmental, we can suppress the resultant measurement error arbitrarily through repetition/feedback \cite{hume_2007}. For high-fidelity operations, we can assume the error from each noise source is additive, allowing us to consider the effect of each error Hamiltonian $\hat{H}_{0}\rightarrow \hat{H}_{0}+\hat{H}_{e}$ separately. If Eq.~(\ref{eq:qnd_def}) is valid at all times, the condition: 
\begin{eqnarray}\label{eq:error_qnd_cond}
    \Big[\hat{H}_{e},\hat{O}\Big] &=& 0,
\end{eqnarray}
will tell us whether or not $\hat{H}_{e}$ preserves QND purity, and, therefore, whether or not we can suppress the resultant error. \\

The measurement sequences discussed in Sec.~\ref{sec:subspace_measurements} comprise four general operations: 1) single qubit rotations on the readout ion 2) application of the geometric phase interaction 3) measurement of the readout ion 4) conditional shelving operations on the logic ion. In Sec.~\ref{sec:subspace} below, we discuss 1-3, which maintain a high degree of QND purity in the presence of the noise mechanisms predicted to dominate the error budget; we also show how to exponentially suppress this effect using classical error correction, which lets us determine the logic ion's subspace with a high-degree of certainty. Operation 4, conditional shelving, clearly violates Eq.~(\ref{eq:qnd_def}). In Sec.~\ref{sec:shelving} we show it is possible, however, to leverage high-fidelity subspace measurements to detect and correct shelving errors\textemdash suppressing infidelities to roughly the same degree we did the subspace measurements.

\subsection{Subspace measurement errors}\label{sec:subspace}

\begin{figure}[b]
\includegraphics[width=0.5\textwidth]{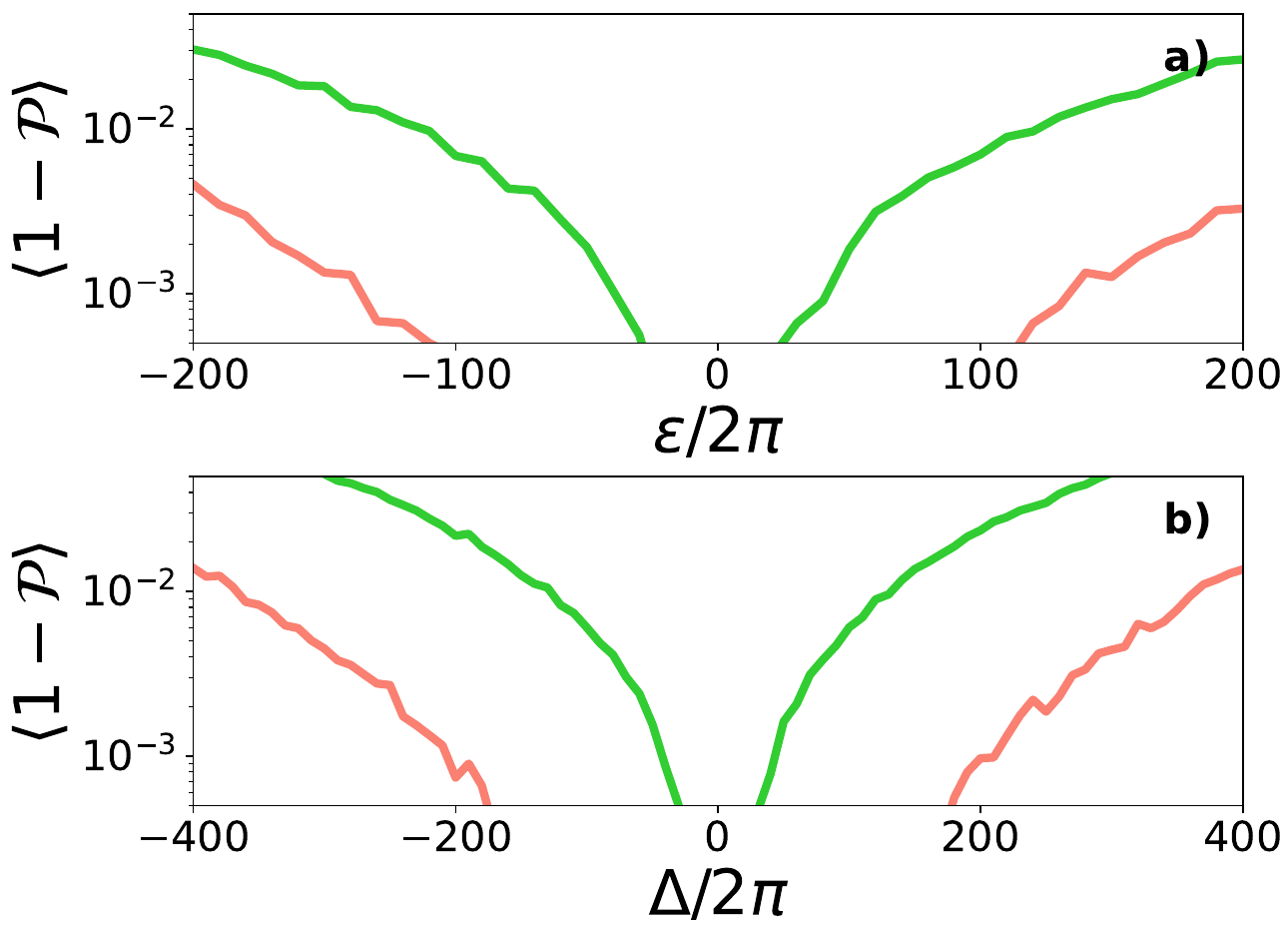}
\centering
\caption{State preparation error $\braket{1-\mathcal{P}}$ for our $^{171}\text{Yb}^{+}$ example in the presence of static a) mode frequency shifts $\varepsilon$, and magnetic field fluctuations  $\Delta$. Here, $\mathcal{P}$ is the probability of correctly identifying the logic ion's state after the operation. Each plot averages a wave function Monte Carlo simulation over $10^{5}$ random trials. We show this without error correction (green top), and with a simple $n=3$ (pink bottom) majority vote code for the two subspace identification steps.}
\label{fig:init_vote}
\end{figure}

To determine logic ion's subspace (see Sec.~\ref{sec:subspace_measurements}), we map its magnetic sensitivity onto the readout ion's state, which we subsequently measure. This process involves only operations applied directly to the readout ion and the geometric phase gate Hamiltonian $\hat{H}_{g}$, both of which commute with our intended observable $\hat{O}\equiv \ket{F,m_{F}}\bra{F,m_{F}}$. If we implement the entangling operation with lasers \cite{leibfried_2003}, however, spontaneous emission will violate QND purity and create a fidelity bottleneck, as it did in Ref.~\cite{hume_2007}. If we generate Eq.~(\ref{eq:gate_pre_rwa}) using a magnetic field gradient, however, spontaneous emission will not be a concern. \\

For $\hat{H}_{e}$ to violate QND purity, it must contain at least one term that does not commute with $\hat{O}$. This immediately excludes any $\hat{H}_{e}$ that does not directly act on the logic ion subspace. Most sources of motional decoherence in geometric phase gates: mode frequency fluctuations, heating, trap anharmonicities \cite{sutherland_2022_1}, Kerr non-linearities \cite{nie_2009}, etc... fall under this category, as do any rotational or measurement errors on the readout ion. The only way $\hat{H}_{e}$ \textit{can} violate Eq.~(\ref{eq:error_qnd_cond}) is by having non-zero off-diagonal matrix elements $\ket{F^{\prime},m_{F}^{\prime}}\bra{F,m_{F}}$ that are large enough to drive population transfer\textemdash not just cause AC shifts. This is why implementing $\hat{H}_{g}$ with a motional frequency magnetic gradient is so appealing \cite{ospelkaus_2008}, since almost every noise mechanism results in phase shift errors, not population transfer errors. It is, therefore, possible we could implement $\hat{H}_{g}$ with relatively noisy fields and rely on classical error correction to reach high measurement fidelities. \\

To provide an example of how to use classical error correction here, we apply a simple order-$n$ (odd) majority vote code to our subspace measurement protocol (see Sec.~\ref{sec:subspace_measurements}). Assume $C$ is some set of logic ion states, and we want to measure if this ion is in subset $A$ or $B$. For an order $n=1$ code, we apply only one logic/readout sequence, determining $A$ or $B$ then proceeding; let $\mathcal{P}_{e,1}\ll 1$ be the probability of a measurement error. To implement an order $n=3$ code, we repeat the logic/readout cycle $3$ times, recording $A$ or $B$ each time, then proceed based on the most frequent outcome:
\begin{eqnarray}
    AAA&\rightarrow & A \nonumber \\
    ABA&\rightarrow & A \nonumber \\
    BAB&\rightarrow & B, \nonumber
\end{eqnarray}
etc... The probability of a measurement error becomes:
\begin{eqnarray}
    \mathcal{P}_{e,3}\simeq 3\mathcal{P}_{e,1}^{2},
\end{eqnarray}
exponentially reducing the probability we determine the subspace incorrectly. This trend continues for higher $n$, where the measurement fidelity decreases $\mathcal{P}_{e,n}\propto P_{e,1}^{(n+1)/2}$. \\

We now apply voting to our $^{171}\text{Yb}^{+}$ initialization procedure, detailed in Sec.~\ref{sec:preparation}. We begin by applying $\hat{U}_{l}(\pi,0)$ then measuring the readout ion $n$ times in a row. If we measured $\ket{0}$ more than $\ket{1}$, we determine the ion is in $\{\ket{0,0},\ket{1,0}\}$, after which we implement the shelving operation $\hat{U}_{s}$ to map $\{\ket{0,0},\ket{1,0}\}\rightarrow \{\ket{1,-1},\ket{1,1}\}$. If we measured $\ket{1}$ more than $\ket{0}$, we determine the ion is already in $\{\ket{1,-1},\ket{1,1}\}$ and we do not apply shelving pulses. We do the same for the second logic readout/sequence, identifying the state as $\ket{1,1}$ if we measured $\ket{1}$ a majority of the shots and $\ket{1,-1}$ if we measured $\ket{0}$. \\

In Fig.~\ref{fig:init_vote}, we use a wave function Monte Carlo algorithm to numerically simulate the probability $\mathcal{P}$ of correctly identifying the state when applying $n=1$ and $n=3$ majority vote codes to our $^{171}\text{Yb}^{+}$ initialization procedure. At the beginning of each trial, we randomly select $1$ of $4$ $S_{1/2}$ Zeeman sublevels of $^{171}\text{Yb}^{+}$. Using $\Omega_{g}/2\pi=5~\text{kHz}$, we numerically integrate Eq.~(\ref{eq:simple_gp_gate}) in the presence of some $\hat{H}_{e}$. To do this, we use static gate mode frequency shifts:
\begin{eqnarray}
    \hat{H}_{e,m}&=&\hbar\varepsilon\hat{a}^{\dagger}\hat{a},
\end{eqnarray}
as shown in Fig.~\ref{fig:init_vote}a, and static Zeeman shifts:
\begin{eqnarray}
    \hat{H}_{e,Z}&=&\hbar\Delta\Big(\hat{J}_{z}+\frac{1}{2}\hat{\sigma}_{z}\Big),
\end{eqnarray}
as shown in Fig.~\ref{fig:init_vote}b. We assume the rest of the gate is ideal, including rotations and measurements on the readout ion. After this, we determine the probability of the readout ion being in $\ket{0}$ and $\ket{1}$. Using these probabilities, we model the measurement process as a quantum jump \cite{molmer_1993} with a $\mathcal{P}_{0(1)}$ chance of measuring $\ket{0(1)}$, after which we apply the projection operator $\ket{0(1)}\bra{0(1)}$ to the wave function and then normalize it. If $n=3$, we repeat this three times, applying $\ket{0,0}\rightarrow\ket{1,1}$ and $\ket{1,0}\rightarrow\ket{0,0}\rightarrow\ket{1,-1}$ when the readout ion's state is $\ket{0}$. The second logic/readout sequence works the same way. At the end of the trial, we determine the probability $\mathcal{P}=|\braket{G|\psi_{f}}|^{2}$ our `guessed' state $\ket{G}$ corresponds to the final wave function $\ket{\psi_{f}}$. In the figure, we stochastically average over $5\times 10^{5}$ randomized trials, plotting the average infidelity $\braket{1-\mathcal{P}}$ versus static mode frequency shifts in Fig.~\ref{fig:init_vote}a and static Zeeman shifts in Fig.~\ref{fig:init_vote}b. Both figures show multiple orders-of-magnitude improvement when going from $n=1$ to $n=3$. It is worth noting that this exponential error suppression makes the process increasingly difficult to model because the necessary number of trials $\rightarrow \infty$ as $\mathcal{P}\rightarrow 1$. That being said, we simulated $n=5$ and, again, saw more than an order-of-magnitude improvement compared to $n=3$; we have excluded these results from the plot because it would be too numerically expensive to make them look nice.

\subsection{Shelving errors}\label{sec:shelving}

\begin{figure}[b]
\includegraphics[width=0.5\textwidth]{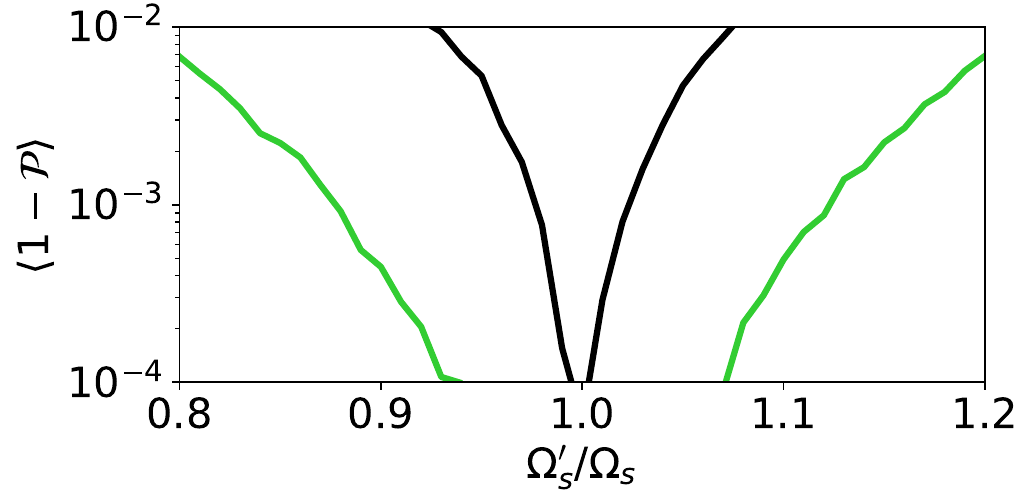}
\centering
\caption{State preparation error $\braket{1-\mathcal{P}}$ versus fractional shift in shelving $\pi$-pulse Rabi frequency $\Omega^{\prime}_{s}/\Omega_{s}$. Here, $\mathcal{P}$ is the probability we have correctly identified the state of the logic ion after stochastically averaging over $10^{5}$ runs (as described in text), $\Omega_{s}$ is the `ideal' Rabi frequency of the shelving pulses, and $\Omega^{\prime}_{s}$ is their actual Rabi frequency. For the black (top) line, we implement the scheme with no additional checks. In the green (bottom) line, we add an additional subspace measurement to flag shelving errors. If a flag is raised, we repeat the shelving pulses.}
\label{fig:shelve}
\end{figure}

To perform SPAM, we have to implement shelving operations on the logic ion that are conditioned on the outcome of the previous measurement(s). Driving population between hyperfine sublevels will, trivially, violate Eq.~(\ref{eq:qnd_def}), so the majority vote code we used to suppress subspace measurement errors does not apply\textemdash directly. We can, however, leverage our ability to measure the logic ion's subspace to determine if we have successfully transferred population from the initial subspace, flagging and correcting errors if necessary. Consider again our $^{171}\text{Yb}^{+}$ state preparation example. If the first measurement sequence returns $\ket{0}$, we know the logic ion is $\{\ket{0,0},\ket{1,0}\}$, so we apply shelving pulses to both states to map them onto $\{\ket{1,-1},\ket{1,1}\}$. The first-order effect of any errors on the shelving pulses will result in residual population in the $\{\ket{0,0},\ket{1,0}\}$ subspace. Since $\{\ket{0,0},\ket{1,0}\}$ and $\{\ket{1,-1},\ket{1,1}\}$ were the first two subspaces we measured, we can simply apply this measurement sequence a second time. If the measurement returns $\ket{0}$, we know the logic ion is in $\{\ket{0,0},\ket{1,0}\}$\textemdash flagging an error\textemdash which we correct by repeating the shelving pulse. Similar to the simulations of the previous section, Fig.~\ref{fig:shelve} shows $\braket{1-\mathcal{P}}$ averaged over $5\times 10^{5}$ trials, randomly selecting from $1$ of the $4$ hyperfine sublevels of $^{171}\text{Yb}^{+}$. As an example noise source, we show the error probability versus the fractional Rabi frequency $\Omega_{s}^{\prime}/\Omega_{s}$, where $\Omega_{s}$ is the `ideal' Rabi frequency. We show this with (green bottom) and without (block top) the additional detection/correction step, again indicating several orders-of-magnitude of improvement.

\section{Conclusion}

In this work, we discussed how we can use the geometric phase gate interaction to map the magnetic sensitivity of a logic ion onto the spin-flip probability of a readout ion. We then discussed how we can use the technique to perform subspace measurements on the ion, which we can use for efficient SPAM and non-destructive leakage detection. Finally, we discussed how to suppress measurement infidelities using error correction.

\section*{Acknowledgments}
I would like to thank R. Matt, R. Srinivas, D. T. C. Allcock, M. Malinowski, and A. S. Sotirova for helpful discussion and comments on the manuscript. I would also like to thank S. D. Erickson for teaching me about QLS. \\

\newpage
\section*{Appendix}

\subsection{$^{137}\text{Ba}^{+}$ preparation example}\label{app:ba_example}
Consider the case where the logic ion is $^{137}\text{Ba}^{+}$, which has $I=3/2$. To perform state preparation, we need to distinguish between all $4I+2=8$ hyperfine sublevels. As it turns out, we can do this in $K=3$ logic/readout sequences, each sequence conditioned on the measurement values before it. Assume we have initialized the readout ion to $\ket{0}$, while the system is in some fully mixed-state of every sublevel $\ket{F,m_{F}}$ in the $S_{1/2}$ ground state manifold. Similar to the $^{171}\text{Yb}^{+}$ example, we set $d\theta = \pi$ and $\phi =0$, mapping the readout ion onto $\ket{0}$ for the even $m_{F}$s, here $A \equiv \{\ket{2,\pm 2}, \ket{2,0},\ket{1,0}\}$, and $\ket{1}$ for the odd $m_{F}$s, here $B\equiv \{\ket{1,\pm 1}, \ket{2,\pm 1}\}$. If measuring the readout ion tells us the logic ion is in $A$, we set $d\theta = \pi/2$ and $\phi_{y}=0$, which leaves the readout ion in $\ket{0}$ if the logic ion is in $AA\equiv \{\ket{2,0},\ket{1,0} \}$ and flips it to $\ket{1}$ if it is in $AB\equiv \{\ket{2,-2},\ket{2,2}\}$. If the first measurement indicates the logic ion is in $B$, then we set $d\theta =\pi/2$ and $\phi=\pi/2$, mapping the readout ion to $\ket{0}$ if the ion is in $BA\equiv \ket{2,1},\ket{1,-1}$ and $\ket{1}$ if it is in $BB\equiv \ket{2,-1},\ket{1,1}$. \\

After the first two logic/readout sequences, the ion will be in $1$ of $4$ possible subspaces, $C_{2}\in \{ AA,AB,BA,BB\}$. If $C_{2}= AA$, we first need to map onto a magnetically sensitive qubit. We can do this by shelving $AA\rightarrow \{\ket{2,1},\ket{1,1}$\}, breaking the near-degeneracies using (for example) dressing fields \cite{sutherland_2024_dressed}. After this, we can apply $\hat{U}_{l}(\pi/2,\pi/2)$, Eq.~(\ref{eq:final_prop_U_l}) in the main text, mapping the readout onto $\ket{0}$ to indicate $AAA = \ket{2,1}$ and onto $\ket{1}$ to indicate $AAB=\ket{1,1}$. If $C_{2} = AB$, we can apply $\hat{U}_{l}(\pi/4,\pi/2)$ which maps the qubit onto $ABA = \ket{2,-2}$ and $ABB = \ket{2,2}$. If $C_{2} = BA$, we can shelve $\ket{2,1}\rightarrow \ket{1,1}$ using a $\pi$-polarized microwave, then apply $\hat{U}_{l}(\pi/2,\pi/2)$ to give $BAA=\ket{1,1}$ or $BAB=\ket{1,-1}$. Finally, if $C_{2} = BB$, we can shelve $\ket{2,-1}\rightarrow \ket{1,1}$, then apply $\hat{U}_{l}(\pi/2,\pi/2)$ which gives $BBA=\ket{1,1}$ or $BBB=\ket{1,-1}$.

\begin{eqnarray} \hat{U}_{l}=\hat{U}_{y,\phi}\hat{U}_{x,\pi/2}^{\dagger}\hat{U}_{g}\hat{U}_{x,\pi/2}
\end{eqnarray}

\bibliography{biblio}
\end{document}